\newcommand{\DYDZ}{$\Delta$$Y$/$\Delta$$Z\,$}        
\newcommand{\Zs}{$Z_\odot\,$}        
\newcommand{\Ms}{$M_\odot\,$}        
\newcommand{\Menvmin}{$M_{env}^{min}\,$}        
\newcommand{\Mpagb}{$M(PAGB)\,$}        
\newcommand{\Ho}{$H_{0}\,$}      
\documentstyle[12pt,aasms4,psfig]{article}
\received{27 July 1998}
\slugcomment{submitted to ApJ}

\lefthead{Yi et al.}
\righthead{UV Upturn as an Age Indicator}
 
\begin{document}

\title{The UV Upturn in Elliptical Galaxies as an Age Indicator}

\author{Sukyoung Yi\altaffilmark{1}, Young-Wook Lee, Jong-Hak Woo}
\affil{Center for Space Astrophysics, Yonsei University, Seoul 120-749, Korea\\ yi@shemesh.gsfc.nasa.gov, ywlee@csa.yonsei.ac.kr, jhwoo@csa.yonsei.ac.kr}

\author{Jang-Hyun Park}
\affil{Center for Astrophysical Sciences, The Johns Hopkins University, Baltimore, MD 21218, and Korea Astronomical Observatory, Taejeon 305-348, Korea\\ jhpark@hanul.issa.re.kr}
 
\author{Pierre Demarque}
\affil{Department of Astronomy, Yale University, P.O. Box 208101, New Haven, CT 06520-8101 \\ demarque@astro.yale.edu}

\author{Augustus Oemler, Jr.}
\affil{Carnegie Observatories, 813 Santa Barbara St., Pasadena, CA 91101 \\ oemler@ociw.edu}


\altaffiltext{1}{present address: NASA/Goddard Space Flight Center, Code 681, Greenbelt, MD 20771}

\begin{abstract}

	The UV upturn phenomenon in elliptical galaxies, although challenging
because of its complexity, is attractive for its potential value as an age 
indicator of old stellar systems.
	This work represents the combined efforts of two population synthesis 
groups with substantially different views to work together to minimize 
uncertainties in modeling and analysis.
	Unfortunately, this study, using the currently available data, 
cannot determine the metallicity of the dominant UV sources, 
one of the most outstanding problems related to the UV upturn phenomenon, 
as some input parameters need to be constrained better.
	We have found, however, that it is feasible to select a more likely
model empirically because different models predict substantially different 
UV-to-$V$ flux ratios as functions of redshift:	metal-rich solutions predict 
a much steeper decline in the UV-to-$V$ flux ratio than metal-poor solutions.
	We show that such differences in model predictions are quite
independent of cosmology and are detectable using current and upcoming space 
UV facilities.

	The various alternatives suggest significantly different ages for
the present epoch giant ellipticals: the metal-rich solutions suggest
30 -- 50\% smaller ages than the metal-poor solutions.
	Thus, an empirical fitting would not only reveal the origin of
the UV upturn but yield independent age estimations for ellipticals.
	We show that this may effectively constrain some of the cosmological 
parameters that predict a unique age for the present epoch galaxies.
	If we use the most recent estimations of \Ho and $\Omega_{0}$, the 
younger, metal-rich models would have no conflict with a cosmology of
a negligibly small $\Lambda_{0}$, while the older, metal-poor models unavoidably 
suggest a substantially large value of $\Lambda_{0}$ (i.e., $\Lambda_{0} \gtrsim 0.63$
for $z_{for}=\infty$) in the context of an inflationary universe.

\end{abstract}

\keywords{galaxies: elliptical and lenticular, cD - galaxies: evolution - galaxies: stellar content - ultraviolet: galaxies, cosmology}

\section{Introduction}

	Various recent studies have suggested that the high ultraviolet (UV) 
fluxes of nearby giant elliptical galaxies (the so-called ``UV upturn'') 
indicate large ages that are comparable with the age of the Milky Way 
(e.g., Bressan, Chiosi, \& Fagotto 1994; Dorman, O'Connell, \& Rood  1995; 
\cite{pl97}; Yi, Demarque, \& Oemler 1998).
	According to these popular theories, the dominant UV sources in
giant elliptical galaxies are low-mass core-helium burning stars, 
such as horizontal-branch (HB) and evolved HB stars that are much more 
populous in older systems.
	If the UV upturn of old systems is truly sensitive to age, as such
population synthesis studies suggest, its value as an independent age 
indicator is extraordinary because spectral evolution in the longer 
wavelength regions (i.e. the visible and infrared) is negligible 
once galaxies are older than a few Gyr. 

	Precise estimates of the ages of both nearby and distant galaxies 
are extremely important to our understanding of the formation and evolution 
of galaxies and of cosmology.
	The significance of accurate age estimates of distant galaxies 
has recently been emphasized by various studies (\cite{dun96}; \cite{spi97}; 
\cite{hea98}; \cite{pea98}), and it is our hope to demonstrate the 
outstanding value of the precise age estimations of nearby galaxies to
cosmology.

	The UV upturn is sensitive to several input parameters that need 
to be further constrained, however.
	While there seems to be a consensus that HB and evolved HB stars are 
most likely the dominant UV sources in giant ellipticals, it is still
debated whether the dominant UV sources are metal-poor 
(\cite{pl97}, hereafter, ``metal-poor HB hypothesis'') or metal-rich 
(\cite{bcf94}; \cite{dor95}; \cite{ydo98}, ``metal-rich HB hypothesis'').
	The problem is that these two scenarios predict significantly
different ages for the same nearby giant ellipticals. 

	According to the metal-rich HB hypothesis, giant ellipticals are 
similar in age to or even younger than old Galactic globular clusters.
	This picture suggests that the metal-rich stars are the most likely 
UV sources in giant ellipticals because, as metallicity increases, 
1) mass loss increases, 2) stars evolve faster (only when metallicity is
markedly higher than the solar), and finally 3) the UV-bright 
helium-burning phase becomes more effective (Horch, Demarque, Pinsonneault 
1992; Dorman, Rood, \& O'Connell 1993; Yi, Demarque, \& Kim 1997).
	On the other hand, the metal-poor HB hypothesis suggests that the 
dominant UV sources are simply much older (and therefore hotter) metal-poor 
HB stars and their post-HB progeny, and, thus, that giant ellipticals are 
about 30\% older than the old halo populations in the Milky Way.
	This large age predicted by the metal-poor HB hypothesis is 
particularly interesting, because it would undermine the effort 
to reconcile the gap between the mean age of old Galactic globular clusters 
estimated via stellar evolution theory and the age of the universe derived
from the measured values of the Hubble constant (\Ho) and the density
parameter ($\Omega_{0}$), if we would not want to invoke a non-zero cosmological 
constant ($\Lambda_{0}$) universe.

	Although weak spectral features found with {\it Hopkins Ultraviolet 
Telescope} seem to suggest a low metallicity ($\approx 0.1$ \Zs, 
\cite{bfdd97}), it is not yet feasible to determine 
the metallicity of the UV sources in giant ellipticals 
spectroscopically, mainly because of the complex nature of the heavy element 
redistribution in the atmospheres of evolved stars 
(e.g., Michaud, Vauclair, \& Vauclair 1983; \cite{heb84}). 
	That is, when stars evolve to become hot HB stars, their atmospheric
chemical compositions significantly deviate from their original compositions 
that dictate their evolution.

	Two different schools (Park \& Lee for the metal-poor HB hypothesis 
and Yi, Demarque, \& Oemler for the metal-rich HB hypothesis) having such 
different views have combined forces in this work to minimize the errors 
both in modeling and in analysis. 
	Our aims are first to attempt to understand the conditions that
create the observed UV upturns and second to use the concept of the UV 
upturn as an age indicator for giant ellipticals.
	First, we have constructed galaxy population models using 
consistent assumptions.
	Then, we predict the evolution of UV-to-$V$ flux ratios as a function
of redshift. 
	A similar effort has already been made by Chiosi, Vallenari, \& 
Bressan (1997) mainly to find differences between the galaxy models whose 
dominant UV sources are either HB stars or PAGB stars.
	However, various studies including their own have already shown that 
PAGB stars are not believed to live long enough to produce the observed amount
of the UV flux in giant ellipticals (see \cite{ct91}).
	Thus, our work is focused on whether the two hypotheses with
different metallicities would lead to different UV spectral evolutionary 
patterns, and if so, whether we would be able to detect them.

	We show that, even if the two hypotheses (metal-poor and 
metal-rich) are nearly equally capable of reproducing the observed 
UV-to-visual flux ratios in present-epoch giant ellipticals, their 
evolutionary predictions are quite different from one another.
	We hope to be able to select the more likely scenario by comparing
various models to future observational data. 
	This will eventually constrain the ages of nearby galaxies 
and thus the epoch of galaxy formation.
	We will also show that such fine age estimations enable us
to determine some cosmological parameters, such as the cosmological
constant, when others are constrained independently. 

\section{Evolutionary Population Synthesis}

	Our independent population synthesis codes generate nearly the same 
results when the same conditions are provided.
	The metal-poor and metal-rich HB hypotheses predict such different 
ages for giant ellipticals mainly because some of their input parameters are 
different.
	As we will see later on, the adopted values of Galactic helium 
enrichment parameter ($\Delta$$Y$/$\Delta$$Z$), the mass loss efficiency, 
and of the chemical evolution model significantly influence the result.
	Thus, it is very important to provide detailed information on what 
values of input parameters and which prescriptions have been used in the 
galaxy modeling (\cite{gr90}; Yi, Demarque, \& Oemler 1997). 

	We have used the evolutionary population synthesis (EPS) technique
to model galaxies. 
	Recent reviews of Yi et al. (1997b) and Park \& Lee (1997)
describe the details of the technique that has been used in this study.
	It is not our intention to discuss the EPS technique in this
paper in detail, and readers should refer to these papers for details.
	Keeping it in mind that it is important to understand the details of 
the modeling process and input assumptions, knowledgeable readers may skip 
this section to \S 3 where we present our favorite models of galaxy evolution.

\subsection{Stellar Libraries}

	For consistency, we have used the stellar evolutionary tracks
constructed using the same input physics, whenever possible.
	They are mainly the Yale group tracks, consisting of the
Yale Isochrones (Demarque et al. 1996) for main sequence (MS) through the 
tip of the red giant branch (RGB) and the post-RGB tracks of Yi, Demarque,
\& Kim (1997). 
	We have adopted Sch\"{o}nberner group's post-asymptotic giant 
branch (PAGB) models (\cite{s79}; \cite{s83}; \cite{bs90}).
	We have used the Kurucz theoretical spectral library (Kurucz 1992) 
and additional spectral models for hot stars constructed using the 
1995 version Hubeny code (Hubeny 1988), as described in Yi et al. (1997b).

\subsection{Mass Loss on the Red Giant Branch}

	The adopted mass loss is extremely important to the integrated UV 
flux. 
	Park \& Lee (1997), a group representative of those using the 
metal-poor HB hypothesis, adopted a fixed amount of mass loss derived from 
the globular cluster HB morphology, assuming the mean age of 13 Gyr for old 
Galactic globular clusters. 
	In order to compensate for the effects of age indirectly, they 
modified the values of the mass loss according to age. 
	The results from such treatments are meant to be close to those 
based on a fixed value of $\eta$, the empirical fitting factor in 
Reimers' mass loss formula (\cite{r75}).
	This treatment, however, ignores the potentially important effects 
of metallicity on the mass loss.
	Most other studies have been based on a fixed value of $\eta$
(Bressan et al. 1994; Yi et al. 1995).
	More recently, Yi et al. (1998) investigated the effects of the 
so-called ``variable-$\eta$ hypothesis'' where $\eta$ increases 
with increasing metallicity, considering the recent hydrodynamic simulations
of the Iowa State group (\cite{bw91}; Willson, Bowen, \& Struck 1996).
	Yi et al. (1997b) clearly demonstrated the large dependence of the
total UV flux on $\eta$.
	However, many EPS studies are still based on values of $\eta$ 
that date back from decades ago, even though many advances have been made in 
stellar astrophysics since then.
	For this reason, we try to find the value of $\eta$ that best matches 
the HB morphology of Galactic globular clusters, using the recent stellar 
evolutionary tracks of the Yale group.

	The Galactic globular clusters located in the inner halo 
(Galactocentric radius $\lesssim$ 8 kpc) are often considered the oldest 
stellar populations in the Milky Way (Lee, Demarque, \& Zinn 
1994)\footnote{Some of the inner halo clusters may have different ages from  
the others. Minniti (1995) suggested that the metal-rich ones with 
Galactocentric radius $\lesssim$ 3 kpc belong to the bulge, and not to
the halo, based on their kinematics, metallicities and spatial 
distributions. However, inclusion or exclusion of these inner-most clusters 
does not alter our results.}.  
	The 45 Galactic globular clusters in the 
inner halo sampled by Lee et al. (1994) have nearly uniform ages, and their 
different HB morphology is mostly due to the difference in metallicity. 
	Therefore, if we can estimate the mean absolute age of the old
Galactic globular clusters, we would be able to determine the best fitting 
value of mass loss by comparing synthetic HB models to the observed HB 
morphology.

	The absolute age of the old Galactic globular clusters, however, 
is not trivial to determine.
	Until recently, the majority of stellar evolutionists have suggested
that they are approximately 15 Gyr old (e.g. Chaboyer, Demarque, \& 
Sarajedini 1996; VandenBerg, Bolte, \& Stetson 1996). 
	However, recent Hipparcos observations of Cepheid variables have
yielded a new zero-point of the Cepheid period-luminosity relation 
(\cite{fc97}; \cite{a97}). 
	Their re-calibration of RR Lyrae absolute magnitudes using the new 
Large Magellanic Cloud and M31 Cepheid distances implies that the mean age 
of the old Galactic globular clusters is approximately 11 Gyr,  
significantly lower than the previous estimates.
	Independently, a set of Hipparcos observations of field MS
stars have led to similarly small age estimates ($\approx$ 12 Gyr) for 
Galactic globular clusters (\cite{r97}; \cite{gra97}; \cite{sw97};
\cite{cha98}). 
	
	Whether the mean age of old Galactic globular clusters is 12 Gyr or
15 Gyr makes a substantial difference to the value of $\eta$ to be adopted.
	If the age of the oldest Galactic globular clusters is younger, 
stars must have experienced more efficient mass loss in order to match the 
observed HB morphology of Galactic globular clusters at smaller ages.
	It is generally believed that the HB morphology at a given metallicity
is a reliable indicator of the mean mass of HB stars.
	For the construction of synthetic color-magnitude diagrams (CMDs), 
we have assumed a truncated Gaussian mass distribution on the HB with a fixed  
mass dispersion parameter, $\sigma$ = 0.04 \Ms, for all metallicities 
(see \S 2.4 for details). 
	We have used the HB type [(B-R)/(B+V+R)] as an HB morphology indicator 
and followed the synthetic HB construction method of Lee, Demarque, \& Zinn 
(1990).

	The best fitting value of mass loss is 0.257 \Ms for the 12 Gyr 
assumption and 0.208 \Ms for the 15 Gyr assumption, as shown in Fig. 1 and
listed in Table 1. 
	Since both values are derived from the old clusters, the value of 
$\eta$ should be obtained using a consistent metallicity.
	If we consider old inner halo clusters as the most metal-poor 
clusters, we might choose $Z=0.0001$. 
	However, when we use the HB morphology fitting method, the clusters
whose HB type values are near 0 are the most influential ones during
the fitting process.
	Therefore, $Z=0.001$ is more appropriate.
	Using newly improved mass loss calculations, the best fitting values 
of $\eta$ are approximately 0.5 and 0.7 for the 15 Gyr and 12 Gyr assumptions,
respectively\footnote{The previous mass loss estimates in Tables 1 -- 4 of 
Yi et al. (1997b) were inaccurate because they used only a small number of
evolutionary tracks for computations. So, we have recomputed the mass loss 
using finer mass grids and a more realistic interpolation routine (Hill 1982). 
Send an email to S.Y. to obtain the tables.}. 
	Iben \& Renzini (1983) pointed out that $\eta$ for 
Galactic globular cluster stars cannot be larger than a critical value.
        If $\eta$ were larger than that, the red giants in Galactic 
globular clusters would have lost all the envelope mass throughout their 
lifetimes in the RGB phase.
	Then, all red giants would have evolved to become PAGB stars
bypassing the HB phase: there would have been no HB stars in Galactic 
globular clusters, which is certainly not true!
	Old Galactic globular clusters were believed to be about 15 Gyrs old,
and with this adopted age, the critical value of $\eta$ was suggested to
be 0.6 by Iben \& Renzini (1983).
	The same would still be true in our study as well, if we were to adopt 
an age of 15 Gyr for old Galactic globular clusters.
	However, in this paper, we have adopted the new, smaller mean age 
for globular clusters, 12 Gyr, which results in a larger estimate of $\eta$
(0.7). 
	Thus, the logic of Iben \& Renzini still holds true as long as 
relative scales are concerned.

	We find the earlier estimation of $\eta$ of Yi et al. (1997b) 
inaccurate, because they assumed that the mass of RR Lyrae stars were
similar to the mean mass of HB stars, even though the mean mass of HB stars 
may be significantly different from the RR Lyrae mass in some circumstances.
	This was why their RR Lyrae-based estimations of $\eta$ were lower 
($\eta = 0.3$ -- 0.5, see their Fig. 3) than the ones based on their 
integrated UV-to-$V$ flux ratio ($\eta = 0.5$ -- 0.7, see their Fig. 15). 
	Consequently, they argued that 
their small values of $\eta$ for metal-poor stars were consistent with those
suggested by the hydrodynamical study of the Iowa State group and suggested 
the so-called variable-$\eta$ hypothesis where $\eta$ increases with 
increasing metallicity. 
	However, if empirical estimates of $\eta$ are approximately 0.5 -- 
0.7, it does not entirely support the variable-$\eta$ hypothesis which suggests
smaller values ($\eta = 0.25$ -- 0.5) for metal-poor stars, although it does 
not rule it out, either. 
	In this study, we have assumed that old Galactic globular clusters are 
about 12 Gyr old on the average and have constructed galaxy models under two 
assumptions of mass loss, i.e., (1) fixed $\eta$ ($=$ 0.7) and 
(2) variable-$\eta$, as shown in Table 2.


\subsection{Minimum Envelope Mass (\Menvmin)}

	Early studies of stellar evolution suggested that HB stars cannot 
have an infinitesimal envelope mass, because the envelopes of 
RGB stars, their progenitors, require a certain minimum mass to exert 
gravitational pressure on the stellar core which triggers the helium core 
flesh at the tip of the RGB (\cite{t70}; \cite{cd80}; 
Cole, Demarque, \& Deupree 1985). 
	While the true value is not yet clear, we have chosen 0.005 \Ms 
as the minimum required envelope mass (\Menvmin) in this study, following 
the recent stellar evolutionary calculations of Sweigart (1998).
	Such a small value of \Menvmin seems required empirically
as well in order to account for the presence of hot HB stars found in several 
globular clusters (\cite{lan96}; \cite{sos97}), assuming that they are  
all products of simple single stellar evolution.

\subsection{Mass Dispersion on the HB: $\sigma$}

	Yi et al. (1997b) have pointed out that the mass loss dispersion 
parameter $\sigma$ (approximately 1.5 times greater than the standard 
deviation $\sigma_{SD}$) also has an important impact on the total UV flux.
	Following Lee (1990), we have adopted $\sigma$ = 0.04 \Ms.
	This value is smaller than the value, 0.06, which was used by 
Yi et al.. 
	This smaller value in the mass dispersion causes a reduction of
the UV flux at a given age until the majority of HB stars become dominantly 
hot.

\subsection{Mass of PAGB stars: \Mpagb}

	The UV light contribution from PAGB stars increases as the assumed 
masses of PAGB stars decrease because their lifetimes correlate inversely
with their masses.
	The mean mass of PAGB stars in old stellar systems with 1 -- 2
\Ms progenitors is believed to be somewhere between 0.55 and 0.60 \Ms  
(Iben \& Renzini 1983; Weidemann \& Koester 1983; Weidemann 1997).
	The luminosities of the PAGB stars found in globular clusters are also 
in agreement with this mass range (between 0.546 and 0.565, de Boer 1987).
	We have therefore assumed 0.565 \Ms as the typical value 
of \Mpagb for the progenies of the MS stars of the mass range 
$M$ = 0.8 -- 1.0\footnote{
	On the other hand, Brown et al. (1998) find that the frequencies 
of PAGB stars in their M31 and M32 samples are in better agreement with a 
larger mass ($\gtrsim 0.633$ \Ms) unless PAGB stars are escaping detection
through a significant amount of extinction.}.
	
\section{Galaxy Models with Realistic Metallicity Distributions}

	For the composite galaxy model construction, we have used the 
metallicity distribution models of Kodama (\cite{ka97}) which are consistent 
with ours in terms of the input physics.
Firstly, his population synthesis models are based on the same stellar
evolutionary tracks as ours. 
	Secondly, Kodama has taken into account the mass dispersion on the HB
in a similar manner to ours.
	Although galactic chemical evolution models are developed generally to
reproduce the ``color-magnitude relation'' of elliptical galaxies
(e.g. \cite{ka97}), many such population synthesis models do not 
take into account a realistic mass dispersion on the HB, one 
of the most important contributors to the UV upturn.
	By and large, inclusion of a mass dispersion on the HB causes
a higher flux in the short wavelength region ($\lesssim$ 4000 \AA).
	Even Kodama's models have not been 
tuned to reproduce the UV part of the spectra of giant ellipticals yet.
	Therefore, the ability and reliability of our composite models 
in matching the observed spectrum in the entire wavelength range are limited. 
	As a result, we have decided not to pay too much attention to detailed 
spectral fittings in this study.

	Whether ``infall'' or ``simple'' (closed box) models for galactic
chemical evolution represent the truth better has been studied extensively 
(e.g. Larson 1972; Audouze \& Tinsley 1976; Chiosi 1980; 
Gibson \& Matteucci 1997), but the answer is still uncertain.
	Since we do not have a preference, we have selected whatever model
generates an integrated spectrum that matches the data better.
	In general, ``infall'' models predict fewer metal-poor
stars and match the near-UV spectrum better when the dominant
UV sources are metal-rich (Yi et al. 1998). 
	On the other hand, ``simple'' models match {\it the observed 
magnitudes} of UV fluxes at relatively smaller ages if the dominant UV sources 
are metal-poor, as shown below through Models A and B. 
	Consequently, we have adopted the ``infall'' (or ``simple'') model in 
model galaxies where the dominant UV sources are metal-rich (or metal-poor). 
	As shown below, the mean metallicity of the dominant UV sources
varies depending on the adopted metallicity distribution.
	In Fig. 2, we display the Kodama metallicity distribution models as 
continuous lines and the integrated distributions used in our population
synthesis as histograms.


	In Table 3, we list four models that are almost equally capable of
reproducing the observed magnitude of the UV upturn in giant ellipticals
(i.e. the ratio between the flux at 1500 \AA\, and that of the Johnson $V$ 
band).
	The magnitudes have been defined as 
$m(\lambda)~=~-2.5~$log~$<\!f(\lambda)\!>$ where $<\!f(\lambda)\!>$ is the 
mean flux in the bandpass. 
	The $<\!f(1500)\!>$ and $<\!f(V)\!>$ are defined by averaging the 
flux within the ranges 1,250 -- 1,850~\AA\, and 5,055 -- 5,945~\AA\,
(\cite{a76}), respectively.
	These magnitudes have been normalized to make Vega's magnitudes 
and colors $V=0.03$, $B-V=-0.01$, $V-R=-0.009$, and $V-I=-0.005$, following
Bessell (1990).
	The UV magnitude at 1,500 \AA, $m(1500)$, has been normalized so as
to our simplified $V$ magnitude becomes identical to the Bessell's $V$
magnitude. 
	As a result, one magnitude difference between $m(1500)$ and $V$
simply means about a 2.5 times difference in the mean flux in these
bandpasses.
	
	Figs. 3 -- 6 provide visual descriptions of these models.
In each figure, the top panel shows the observed spectrum of one of the
UV-strong galaxies, NGC\,4552, and a model of a particular age that matches the
observed spectrum.
	NGC\,4552 is the only galaxy whose entire spectrum was available
at the time of this study.
	Besides, we have chosen it because we wanted to estimate the age of 
one of the oldest giant ellipticals in order to discuss cosmological 
implications, as our models suggest that the stronger the UV flux, the 
older the galaxy\footnote{This excludes galaxies with recent star formation.}.
	Sources of the composite spectrum of NGC\,4552 are
(1) $\leq$ 1800 \AA: {\it Hopkins ultraviolet Telescope} spectrum of NGC\,4552 
(\cite{bfd95}), (2) 1800 -- 3300 \AA: mean $IUE$ spectrum of UV-strong galaxies
(\cite{b88}), (3) 3300 -- 3700 \AA: UV-strong galaxy NGC\,4649 (\cite{a96}), 
and (4) $\geq$ 3700 \AA:  average of Bica's E1 group galaxies (\cite{bica88}).
	The middle panel shows the light contribution from
the stars in various evolutionary stages, and the bottom panel shows
the light contribution from different metallicity groups of stars.
	We summarize the results of our analysis of these models briefly.


	Model A: This is the only model in Table 3 whose dominant UV sources 
are metal-poor HB stars in the central helium burning phase.
	We define the central helium burning phase as a part of the
core helium burning phase that extends from the zero age HB to the point
where helium is almost exhausted in the stellar center ($Y_{c} = 0.01$),
following Yi et al. (1997b, see their Fig. 13 for illustration). 
	Similarly, the shell helium burning phase covers the rest of the
core helium burning phase which is more evolved and generally
short-lived but much more luminous than the central helium burning phase.
	Input parameters have been chosen to maximize UV light production
from metal-poor stars for the purpose of demonstrating the  conditions under 
which metal-poor stars become the major UV sources. 
	These parameters are all acceptable under current observational 
constraints.
	For example, we have used the Kodama-``simple'' metallicity 
distribution model which predicts more than twice as many metal-poor stars as 
his infall model does. 
	In addition, we have adopted a slightly smaller value of \DYDZ 
($= 2.0$) than generally accepted (\DYDZ $\approx 2.5$, Peimbert [1995]).
	Under this configuration, nearby UV-strong giant ellipticals, such
as NGC\,4552, need to be as old as 15.4 Gyr, 28\% older than old Galactic 
globular clusters, in order to match the observed strength of the UV upturn 
(see also Park \& Lee 1997).

        Although \DYDZ = 2.5 (\cite{p95}) has been popular in various studies 
(e.g., \cite{bcf94}), its true value is still somewhat uncertain: some studies 
suggest values as high as 3 -- 6 (\cite{l79}; \cite{m92}; \cite{pag92}) while
others suggest values as low as 2.0 
(Izotov, Thuan, \& Lipovetsky 1997)\footnote{Izotov et al. (1997)'s smaller 
estimate of \DYDZ does not necessarily mean a smaller UV light production
of metal-rich stars, however, because it is a result of a larger 
primordial helium abundance ($Y_{P}$ = 0.245) rather than that of a 
smaller present epoch helium abundance. In fact, \DYDZ = 2 with $Y_{P} = 0.245$
suggests $Y = 0.325$ for $Z = 0.04$ which is close to the helium abundance
($Y = 0.33$) of \DYDZ = 3 with $Y_{P} = 0.23$. Since it is $Y$, not \DYDZ, 
that causes metal-rich populations to be UV bright, their discovery does not 
change the conclusion from our analysis.}.
        Guenther and Demarque (1997), through their
stellar evolution study, recently suggested that the estimated initial 
chemical composition of the Sun is ($Y$, $Z$) = (0.2741, 0.020), which would 
then suggest \DYDZ $\approx$ 2.2 when the primordial composition is assumed to 
be ($Y$, $Z$) = (0.23, 0.00).


	Model B: This model is a twin of Model A. The only difference is
that it is based on the Kodama-``infall'' model. Now, the model galaxy 
has many fewer metal-poor stars than the ``simple'' model, so at the
same age (15.4 Gyr), it does not have enough UV light to match the data.
However, it quickly develops additional hot HB stars in its metal-rich 
populations and finally matches the data at 15.9 Gyr. 
	As one might notice here, the pace of developing hot stars is much 
faster in metal-rich populations.
	We will scrutinize this effect in \S 4.1.
	In this model, metal-poor stars still account for a substantial 
40\% of the total UV light.
	It is interesting that the predicted nature of the UV sources 
is so sensitive to the adopted metallicity distribution.
	

	Model C: This model has been constructed with the 
conventional values of \DYDZ ($= 2.5$) and $\eta = 0.7$.
	We have adopted the ``infall'' model in order to allow a larger
fraction of metal-rich stars so that the model matches the data somewhat
better in the near UV.
	The main UV light emitters are highly metal-rich ($>$ \Zs) evolved HB 
stars, as the UV-bright core-helium burning phase becomes more significant 
with increasing \DYDZ.
	According to this model, nearby giant ellipticals that show prominent 
UV fluxes are nearly as old as old Galactic globular clusters (10.7 Gyr) 
but not necessarily older. 

	This may have a significant implication for the estimation of
the true \DYDZ.
	A larger value of \DYDZ is one of the two main
driving forces that make metal-rich stars the dominant UV sources in these
models (the other being the larger mass loss when Reimers' formula with 
a fixed $\eta$ is used [Yi et al. 1998]).
	If the true value of \DYDZ for metal-rich ($>$ \Zs) stars is 
substantially smaller than 2.5, our EPS models with a choice of the most
natural parameters would advocate metal-poor HB stars as the dominant
UV sources.
	Conversely, if studies prove that the dominant UV 
sources are metal-poor, one may regard it as an indirect evidence for a 
small \DYDZ\, in metal-rich ($>$ \Zs) stars in giant elliptical galaxies.


	Model D: This model is based on the variable-$\eta$ hypothesis.
	As we allow a larger value of $\eta$ for a more metal-rich population, 
it obviously generates a larger UV flux at a given age.
	For this reason, this model matches the data at a very small age 
(7.6 Gyr).
	According to this model, even the oldest giant ellipticals in the 
present epoch are only 60 -- 70\% as old as old Galactic globular clusters. 
	This prediction may be in conflict with some recent observations
(e.g. Bower, Lucey, \& Ellis 1992; Spinrad et al. 1997; Kodama \& Arimoto 
1997).

	All these models have been selected by eye fitting. 
	These models are less than perfect in matching the data, especially 
in the near UV and in the infrared.
	However, we should remind ourselves that our models are based on 
theoretical spectral libraries (mainly, Kurucz 1992) which are known to have
some limitations in the UV and in the infrared (e.g. Morossi et al. 1993; 
Bessell, Castelli, \& Plez 1998; Heap et al. 1998).
	Additional limitations come from our poor understanding of
the temperature distribution of HB stars in Galactic globular clusters.
	For example, such details as the spectral slope in the 
UV flux are affected by possible additional sources of mass 
loss that may take place on the HB or during binary evolution, neither of 
which are taken into account in this study.
	Such processes would cause a more strongly bimodal temperature
distribution of HB stars than our simple, single-Gaussian model predicts, 
causing
the far-UV spectrum to be steeper and the near-UV spectrum to be lower, both
of which are desired for a better matching to the observed spectrum.

	Keeping all these uncertainties in mind, we conclude that all models
in Table 3 are nearly equally capable of matching the spectrum of NGC\,4552.
	Multiple solutions exist to the UV upturn phenomenon, and we
cannot determine the mean metallicity of the UV sources in giant ellipticals
by matching the present epoch spectrum alone.
	However, we will show in the next sections that the large difference
in the age of ellipticals predicted by different models allows us to 
answer this question if we use the evolution of UV-to-$V$ flux ratios as
functions of redshift as a new tool.

\section{Photometric Evolution of the UV Upturn}

\subsection{The UV Upturn as a Function of Time}

	Hot HB stars take long to develop, and, for this reason, older 
galaxies show stronger UV fluxes no matter what the true mean metallicity of 
the dominant UV sources is.
	This effect is shown in Fig. 7 where the UV-to-$V$ flux ratios are
shown as functions of age for the models described in \S 3.
	Once stars are older than a few Gyr, the UV-to-$V$ flux ratio 
increases monotonically with time and thus can be used as an age indicator.

	All models have qualitatively the same trend in the manner of 
developing a strong UV flux. 
	At small ages ($\lesssim$ 5 Gyr), the UV-to-$V$ flux ratio decreases
with time because MS stars, then the dominant UV sources, get
cooler with time.
	Thereafter, the UV-to-$V$ flux ratio steadily
increases with time, first as a larger number of MS stars develop
into PAGB stars, and next as hot HB stars slowly but gradually develop.
	This ``onset of the UV upturn'' has been pointed out earlier
(\cite{bcf94}; \cite{tan96}; \cite{ydo98}).

	During the first few Gyr of this rise, PAGB stars dominate
the UV spectrum, but the total UV flux is still very low.
	Hot HB stars gradually take over, however, as a galaxy gets
older and develops low-mass HB stars.
	The exact timing of the transition between these two stages 
(``the PAGB epoch'' and ``the HB epoch'') depends on metallicity, as
pointed out by Yi et al. (1998).
	Despite the similarity in the trends of UV flux development, 
different models infer ages of giant ellipticals that are different by a 
factor of two (between 8 and 16 Gyr)!
	

	Note that the pace of the UV upturn development is faster in Models C 
and D which predict relatively smaller ages for giant ellipticals than the 
others.
	For example, $m(1500) - V$ rises from 4.0 to 2.0 (more than 6 times 
in linear scale) in less than 2 Gyrs in Models C and D, whereas it takes 
nearly 8 Gyrs (a half of their ages) in older models.
	This is mainly because the production (or appearance) of an appreciable
number of hot HB stars is much more abrupt in metal-rich populations.
	This effect can be understood more directly by looking at the HB
evolutionary tracks, as illustrated in Fig. 8.
	In the metal-poor case (top panel of Fig. 8), HB stars gradually 
become hotter as their masses become smaller, while it is much more abrupt in 
the metal-rich case.
	Since their masses become smaller monotonically as the population 
becomes older, HB stars become hotter smoothly as the
population ages: the ``age-HB temperature relation''.
	However, the ``age-HB temperature relation'' is much more abrupt in 
the metal-rich case (bottom panel), mainly because of the large opacity 
effects.
	That is, until HB stars become very light in mass the large opacity
effects in metal-rich stars dominate the age effect and prevent them from 
becoming hot.
	This difference in the pace of the UV upturn development can be
detected if we obtain UV-to-visual colors as functions of redshift, as
described in the following section.
	
\subsection{The UV Upturn as a Function of Redshift}

	The importance of the UV spectral evolution in elliptical galaxies
to cosmology has been addressed by various authors, but most notably by
Guiderdoni \& Rocca-Volmerange (1987).
	In contrast to the longer wavelength regions, the signature and amount 
of the spectral evolution can be readily found in the UV even at large ages.


	Fig. 9 displays the evolution models of the UV-to-$V$ flux ratio
as a function of lookback time, another way of viewing Fig. 7.
	While all four models reproduce the present observed flux ratio
successfully, their evolutionary paths are quite different.
	Note in this diagram that we do not have to look too far in the
past to select the most likely model: only a few Gyr of lookback time 
is enough. 


	We can construct models in totally observable quantities.
	Fig. 10 displays the UV-to-$V$ flux ratio as a function of 
redshift for the models described in \S 3.
	The lookback time and luminosity distance have been computed using
the expressions shown in Carroll, Press, \& Turner (1992).
	First, we have adopted \Ho $= 64$ from Riess, Press, \& Kirshner 
(1996), and $\Omega_{0} = 0.15$ from (Trimble 1987) in this paper.
	Then, in order to accommodate the quite old models (Models A \& B), 
we have arbitrarily chosen a large value of the cosmological constant
($\Lambda_{0}$ = 0.85).
	The estimated ages of UV-strong nearby giant ellipticals are listed 
in Table 3.
	To present these models all under the same particular set of
parameters, different values of the epoch of galaxy formation $z_{for}$ 
have been used for different models, as shown in Fig. 10.
	We do not intend to advocate any of these adopted cosmological 
parameters in this paper, as they have been chosen only for illustrative
purposes.


	The second panel in Fig. 10 shows $m(1500)$ as a function of redshift.
	All models in Fig. 10 are in observed magnitudes, neither 
redshift-corrected nor evolution-corrected.
	The Models A and B that predict larger ages for nearby giant 
ellipticals are brighter than the others by 1 -- 2 magnitudes in the range of
$z = 0.1$ -- 0.4, because their UV flux has been relatively slowly 
increasing with time to reach the current level, as shown in Figs. 7 \& 9.
	This effect is much more obvious in the color vs redshift frame
(bottom panel in Fig. 10).
	The older models, A and B, exhibit a nearly steady increase in the
flux ratio throughout a wide range of redshift, whereas younger models, C
and D, show a recent dramatic increase in the flux ratio.
	We believe that such a large difference in color (about 1 -- 2
magnitudes at $z$ = 0.1 -- 0.4) between different models should be readily 
detectable, using present and upcoming UV space probes, such as $STIS$ on the 
{\it Hubble Space Telescope} and {\it GALEX} (Martin et al. 1998).

	Such a large difference in $m(1500) - V$ between ``young'' 
(relative to the Milky Way)
and ``old'' galaxy solutions is not sensitive to the adopted cosmology.
	For instance, Fig. 11 shows Models A and C as two representative 
cases.
	Cosmological parameters have again been chosen for demonstration 
purpose only.
	We have chosen $z_{for} = 4.3$ so that Model A not only represents 
an ``old'' solution to the UV upturn problem but also satisfies a solid 
prediction of the Inflation model 
($\Omega_{total} \equiv \Omega_{0} + \Lambda_{0} = 1$, Guth [1981]). 
	Remember that conventional inflation models predict
a flat ($\Omega_{0} + \Lambda_{0}$ = 1) universe.
	Model A in Fig. 11 is identical to Model A in Fig. 10.
	However, Model C differs from the Model C in Fig. 10 in terms of 
$\Lambda_{0}$ and of $z_{for}$.
	Despite the change in cosmology, we expect to find the (qualitatively) 
same difference between Model A and Model C.
	Thus, the indications of the age of old giant ellipticals can be found 
through this technique regardless of the details in cosmology.


	Once the age is determined, we can constrain a cosmological 
parameter, provided that other parameters can be independently constrained, 
because the age of the oldest population is uniquely determined by only a 
few parameters (\Ho, $\Omega_{0}$, $\Lambda_{0}$, $z_{for}$). 
	As shown in Fig. 11, the data that match Model A (``old'' model)
would definitely support a substantially large value of $\Lambda_{0}$ and 
$z_{for}$ (e.g., $\Lambda_{0} \gtrsim 0.7$ for $z_{for} \lesssim \infty$).
	Otherwise, it would not be possible for a universe of \Ho = 64
and of $\Omega_{0}$ = 0.15 to contain such an old galaxy in the first place.
	This may appear radical.
	For instance, through graviational lensing studies, Kochanek (1996)
found $\Lambda_{0} \lesssim 0.66$ at 95\% confidence in flat cosmologies.
        It is, however, interesting to note that some of the recent supernova 
observations have indicated similarly large values of $\Lambda_{0}$ 
(e.g., Riess et al. 1998). 
	Younger models (Models C \& D), on the contrary, would not be in
conflict with a negligible-$\Lambda_{0}$ universe.
	Such conclusions are obviously subject to uncertainties in 
the adopted values of \Ho and $\Omega_{0}$.

\section{Summary and Conclusions}

	Two population synthesis groups that have had different views 
regarding the origin of the UV upturn have here worked together to reduce 
some sources of uncertainties in modeling and analysis.
	We believe that our new models  in this study
are more reliable than our previous ones.
	Despite such efforts, whether the dominant UV sources are metal-poor 
or metal-rich, the most outstanding question regarding the origin of the 
UV upturn, cannot be answered directly using the spectra of present epoch 
galaxies alone.
	This is mainly because several input parameters (e.g. \DYDZ, $\eta$, 
and the metallicity distribution inside galaxies) affect the UV flux in 
degenerate manners, while they need to be constrained better.
	However, we can choose a more likely model empirically.

	Different models predict different evolutionary paces of the
UV flux development.
	When the mass loss is assumed to follow the empirical formula of
Reimers (1975) with a fixed efficiency, the models with a relatively larger
value of \DYDZ ($= 2.5$) suggest that the dominant UV source in ellipticals 
are metal-rich and giant ellipticals are younger than or similar in age to 
old Galactic globular clusters.
	They show a steep decline in $m(1500)-V$ flux ratio 
with increasing redshift in the range of $z =$ 0.0 -- 0.2.
	If a relatively smaller value of \DYDZ ($= 2.0$) is used,
models predict that metal-poor stars are at least as important as metal-rich 
stars as UV sources.
	Giant ellipticals are predicted to be 30\% older than the Milky Way in
this scenario.
	When the mass loss efficiency is assumed to correlate positively with 
metallicity (the variable-$\eta$ hypothesis), the UV
sources in giant ellipticals are suggested to be dominantly metal-rich.
	Since metal-rich populations generally develop hot HB stars much more
abruptly than metal-poor populations, these models predict 
a rapid decline in the UV-to-$V$ flux ratio with increasing redshift.
	In this scenario, nearby giant ellipticals are only 60 -- 70\% 
as old as the Milky Way.
	
	The difference in $m(1500)-V$ between these models appear
to be large enough for detection using current and upcoming space facilities,
such as $STIS$ on the $HST$ and $GALEX$.
	This empirical fitting will not only help us select a more likely
solution over others but also provide an important clue to the mean age of 
giant elliptical galaxies.
	This applicability of the UV upturn as an independent age indicator 
is extraordinary because no other obvious photometrically-selected age 
indicators exist for ``old populations'' yet.
	We have also shown that such model predictions are quite independent
of cosmology.

	It is important to determine the ages of giant ellipticals
because they are often suspected to be the oldest populations in the 
universe and the ages of the oldest populations constrain cosmology.
	We have demonstrated that the age-sensitivity of the UV upturn may
effectively constrain one of a set of cosmological 
parameters that predict a unique age for the present epoch galaxies.
	If we use recent popular measurements of \Ho ($= 64$) and
$\Omega_{0}$ ($=0.15$), the older, metal-poor models unavoidably support a universe
with a large value of $\Lambda_{0}$ ($\gtrsim 0.85$ for $z_{for} = 4.3$ or
$\gtrsim 0.63$ for $z_{for} = \infty$), whereas the younger,
metal-rich models would have no conflict with a universe with a
negligibly small value of $\Lambda_{0}$.

	A caveat in this proposed observational test is a possible 
contamination from episodic star formation in giant Es at $z = 0$ -- 1.

\acknowledgements

	We are grateful to Taddy Kodama for providing his metallicity 
distribution models. We also would like to thank Richard Larson and Sydney 
Barnes for constructive comments.
	This work was supported by the Creative Research Initiative Program 
of the Korean Ministry of Science \& Technology (Y.-W.L., S.Y.).
	Part of this work was performed while S.Y. held a National Research
Council-(NASA Goddard Space Flight Center) Research Associateship.
	J.-H.P thanks Korea Science and Engineering Foundation for 
Postdoctoral Fellowship and Arthur Davidsen for supporting his stay in 
Center for Astrophysical Science, the Johns Hopkins University.

{}

\newpage

\figcaption[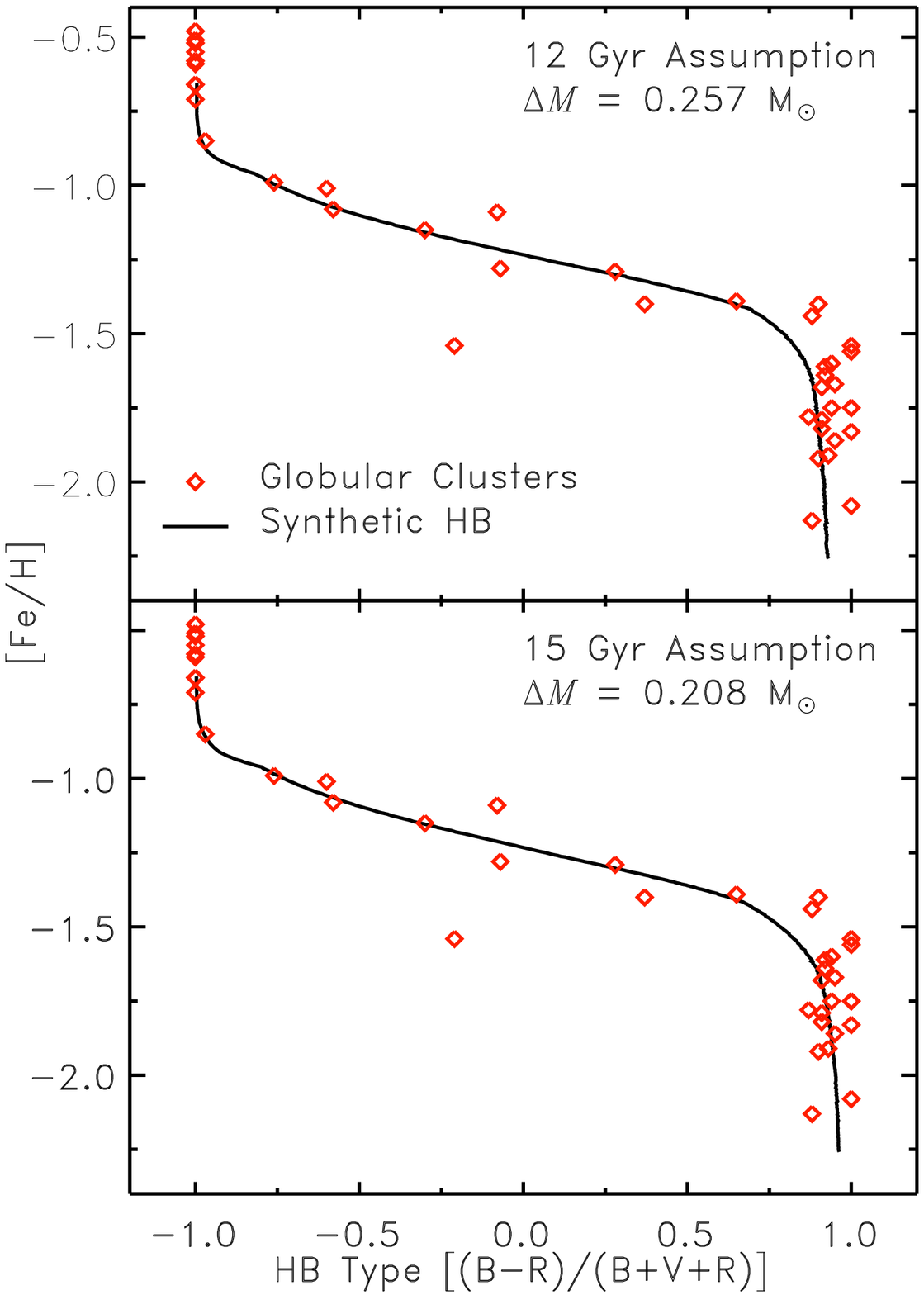]{
The estimation of $\eta$ is based on the synthetic model fitting to the 
HB morphology of old Galactic globular clusters. The fitting suggests that
the amount of mass loss at $Z$ = 0.001 during the red giant phase is 0.257~\Ms
if clusters are assumed to be 12 Gyrs old or 0.208 \Ms if 15 Gyrs old, 
respectively.
\label{fig1}}

\figcaption[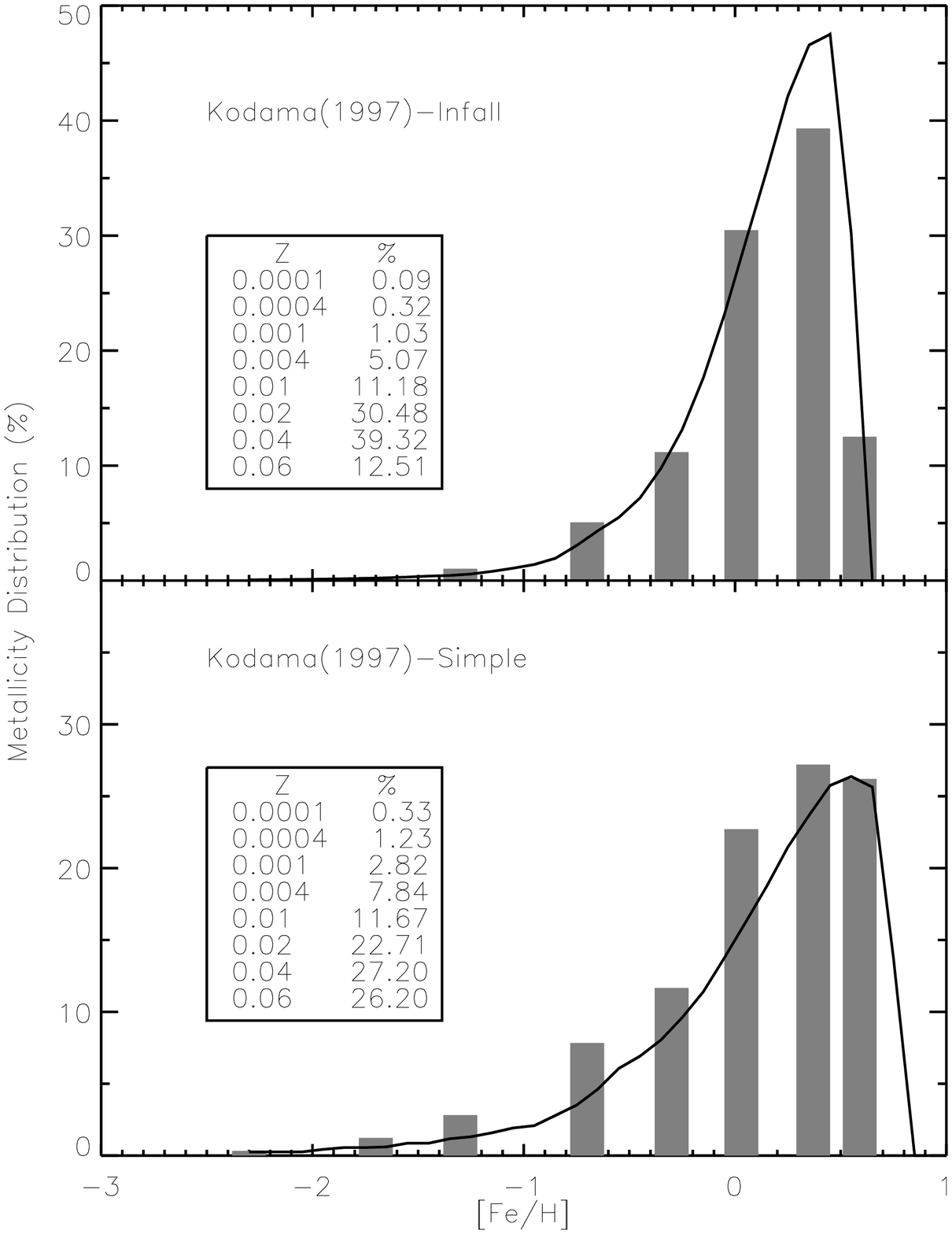]{
{\it Solid line:} the adopted model metallicity distribution from Kodama 
(1997). {\it Histogram and table:} integrated values for our metallicity grids.
\label{fig2}}

\figcaption[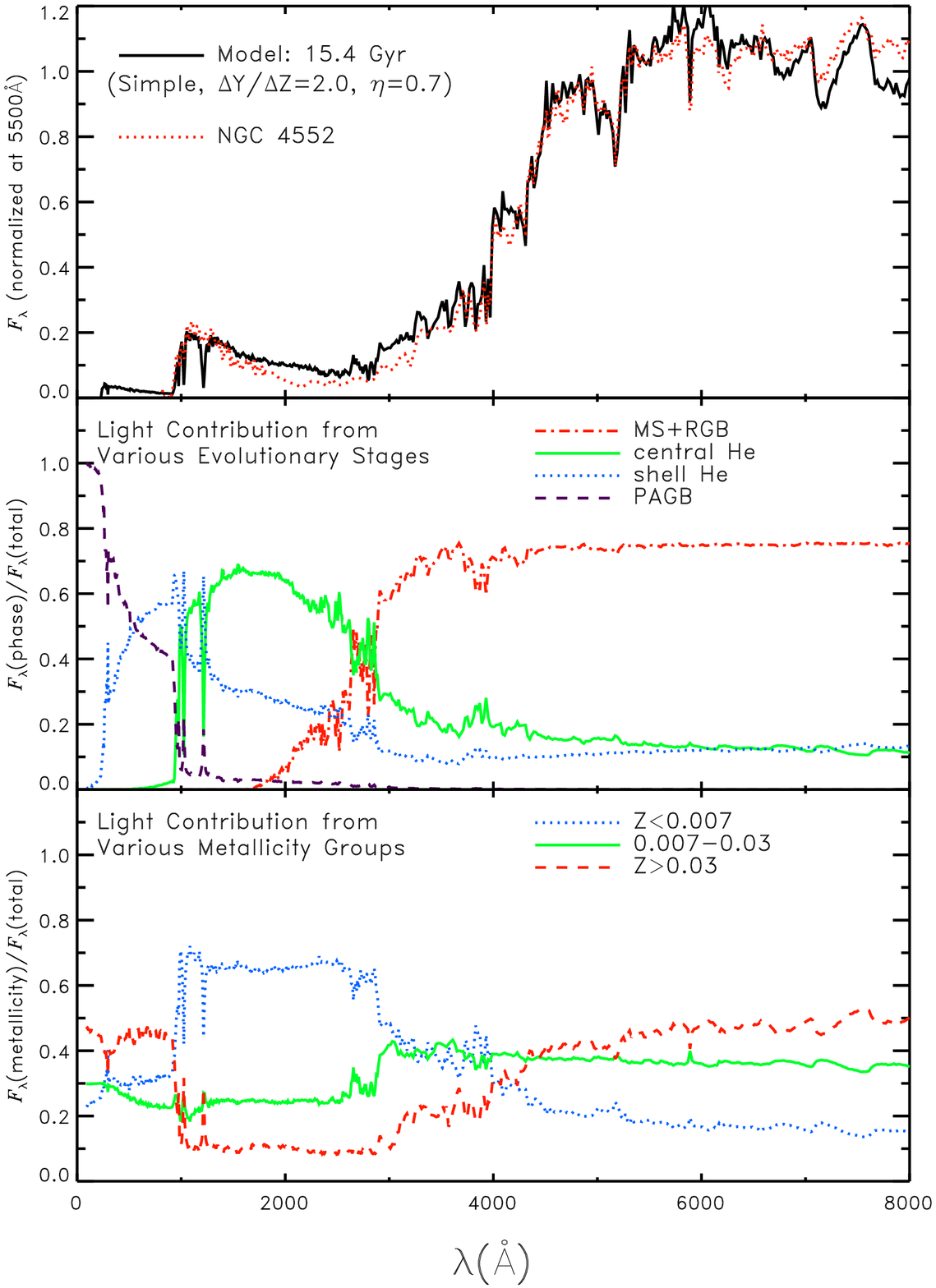]{
Visual description of Model A in Table 3. 
{\it Top panel}: comparison between the observed spectrum and a model.
{\it Middle panel}: the light contribution from various evolutionary stages.
The central helium burning phase is defined as the less evolved part of the 
core helium burning phase (central helium abundance larger than 0.01, see 
text), and the shell helium burning phase is the rest of the core helium 
burning phase which includes such UV bright phases as AGB-manqu\'{e} phase
(Greggio \& Renzini 1990) and the ``slow blue phase'' (Horch et al. 1992).
{\it Bottom panel}: the light contribution from various metallicity groups.
\label{fig3}}

\figcaption[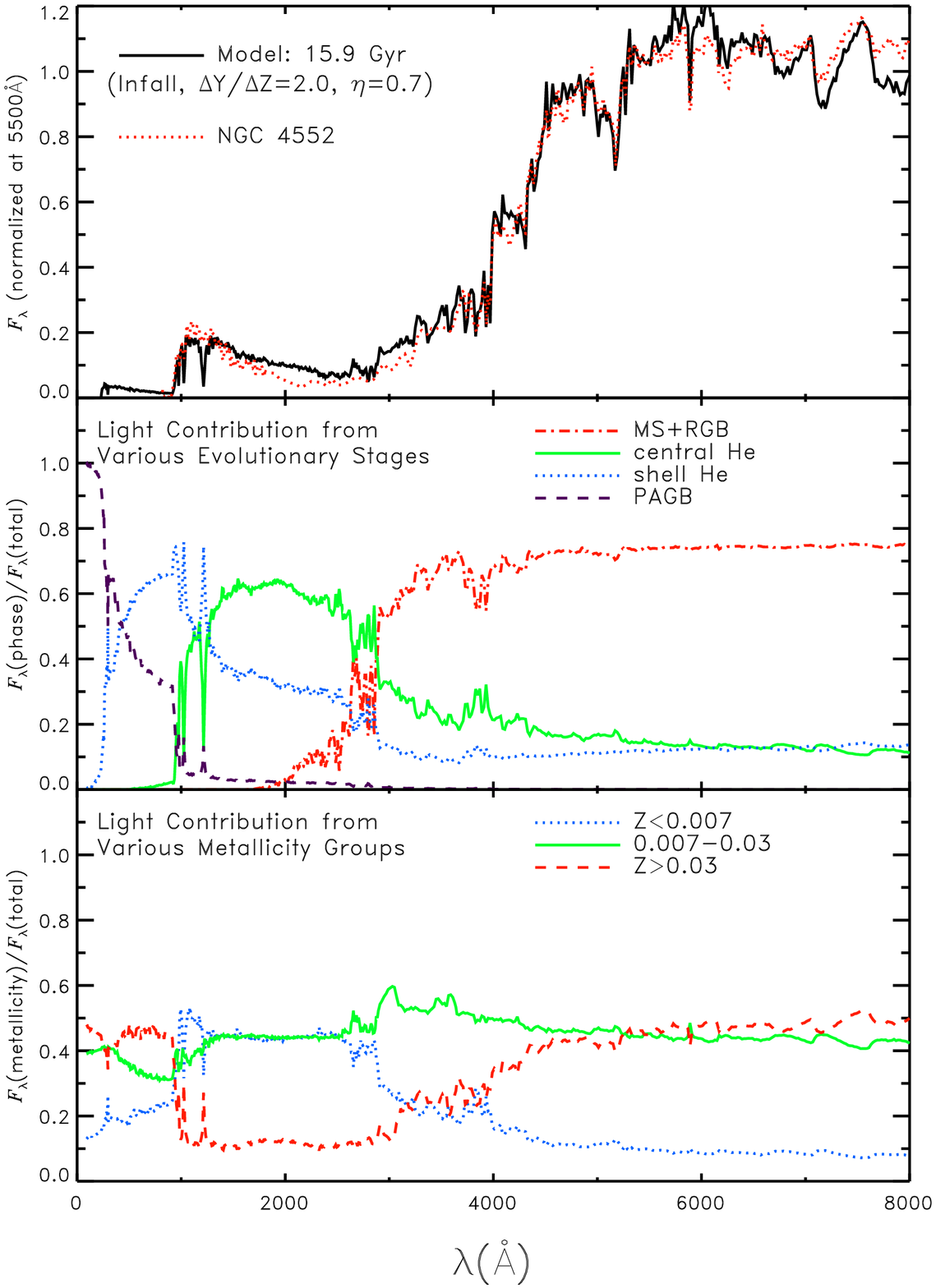]{
Same as Fig. 3, but for Model B.
\label{fig4}}

\figcaption[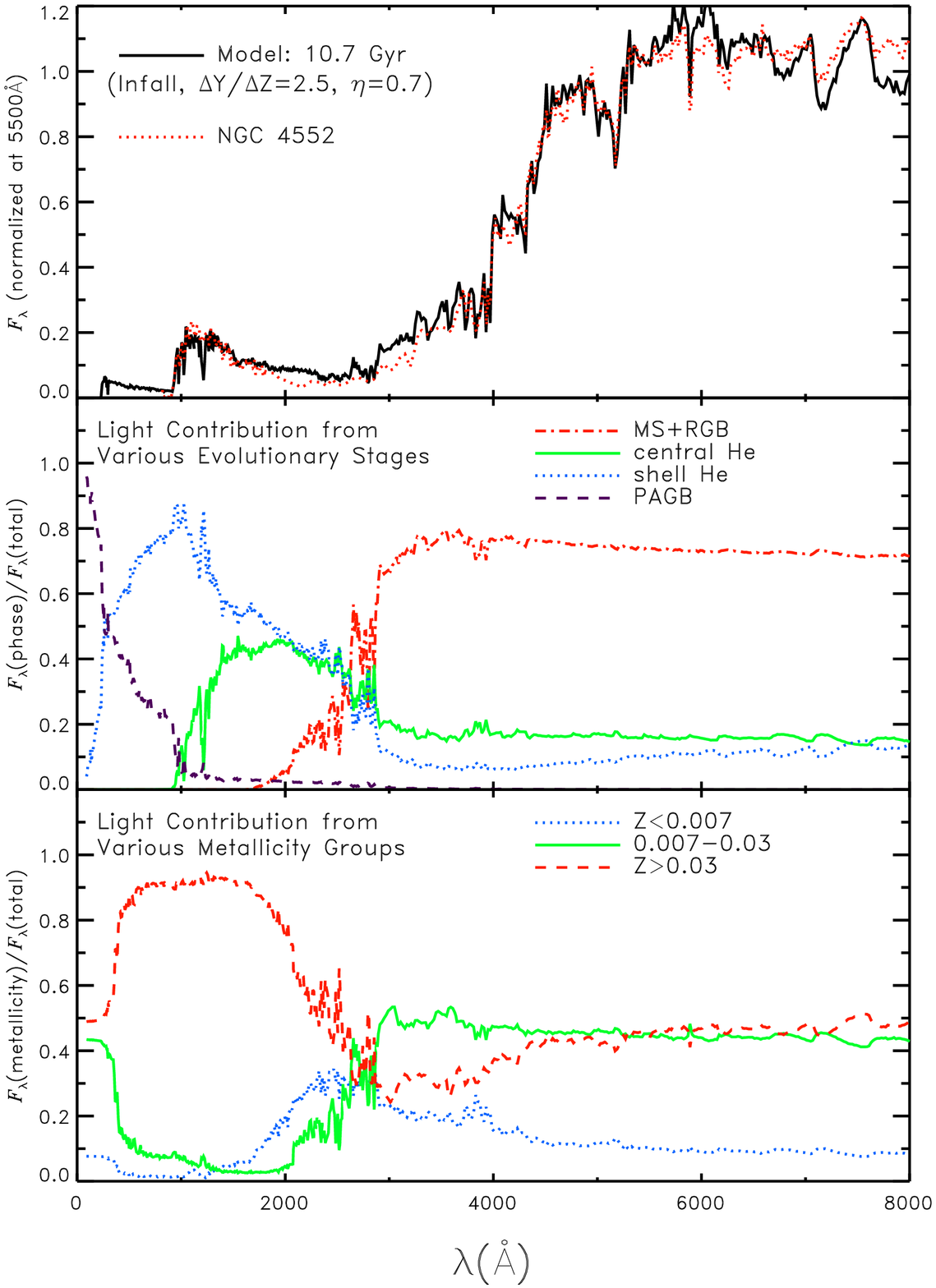]{
Same as Fig. 3, but for Model C.
\label{fig5}}

\figcaption[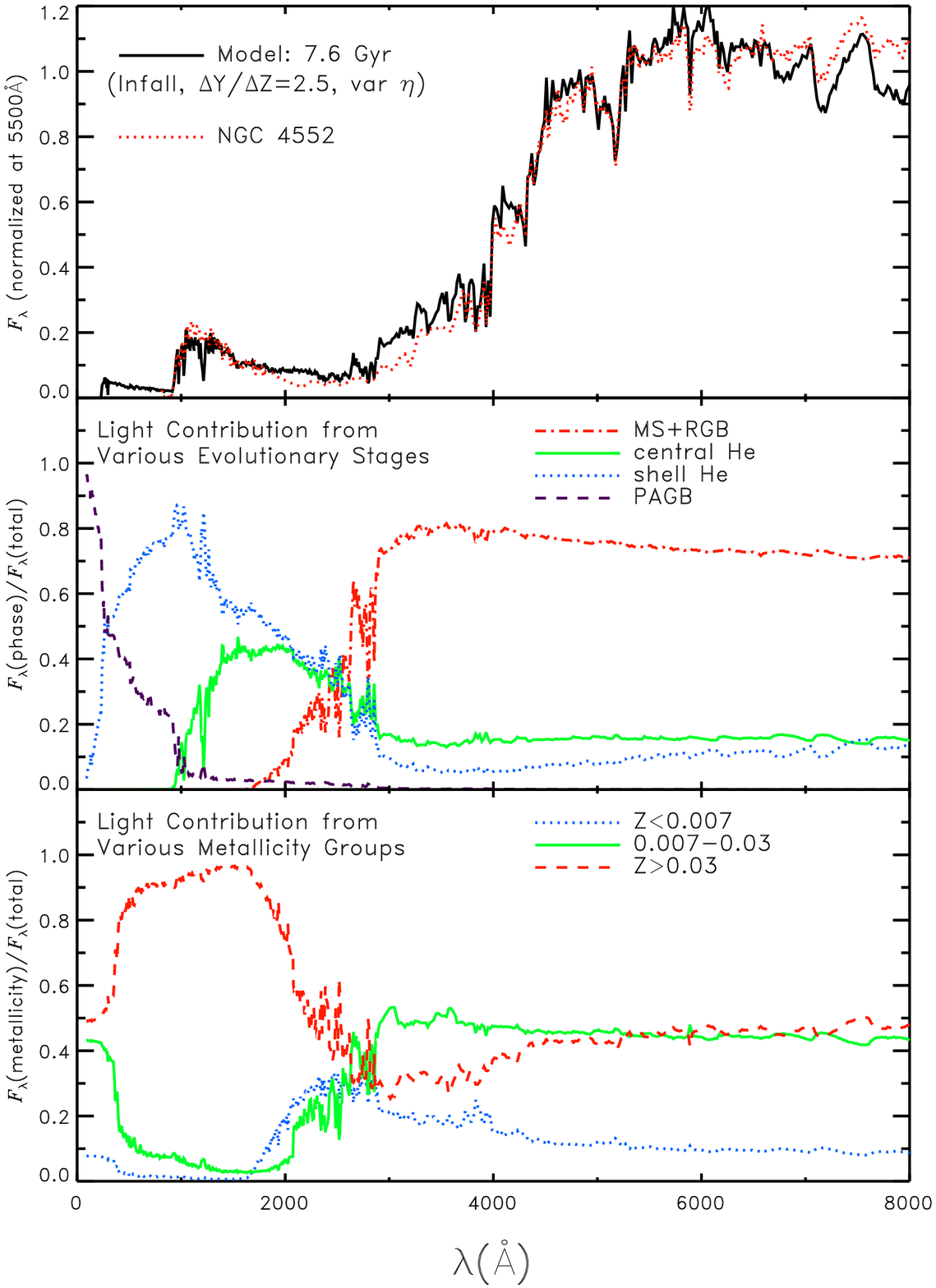]{
Same as Fig. 3, but for Model D.
\label{fig6}}

\figcaption[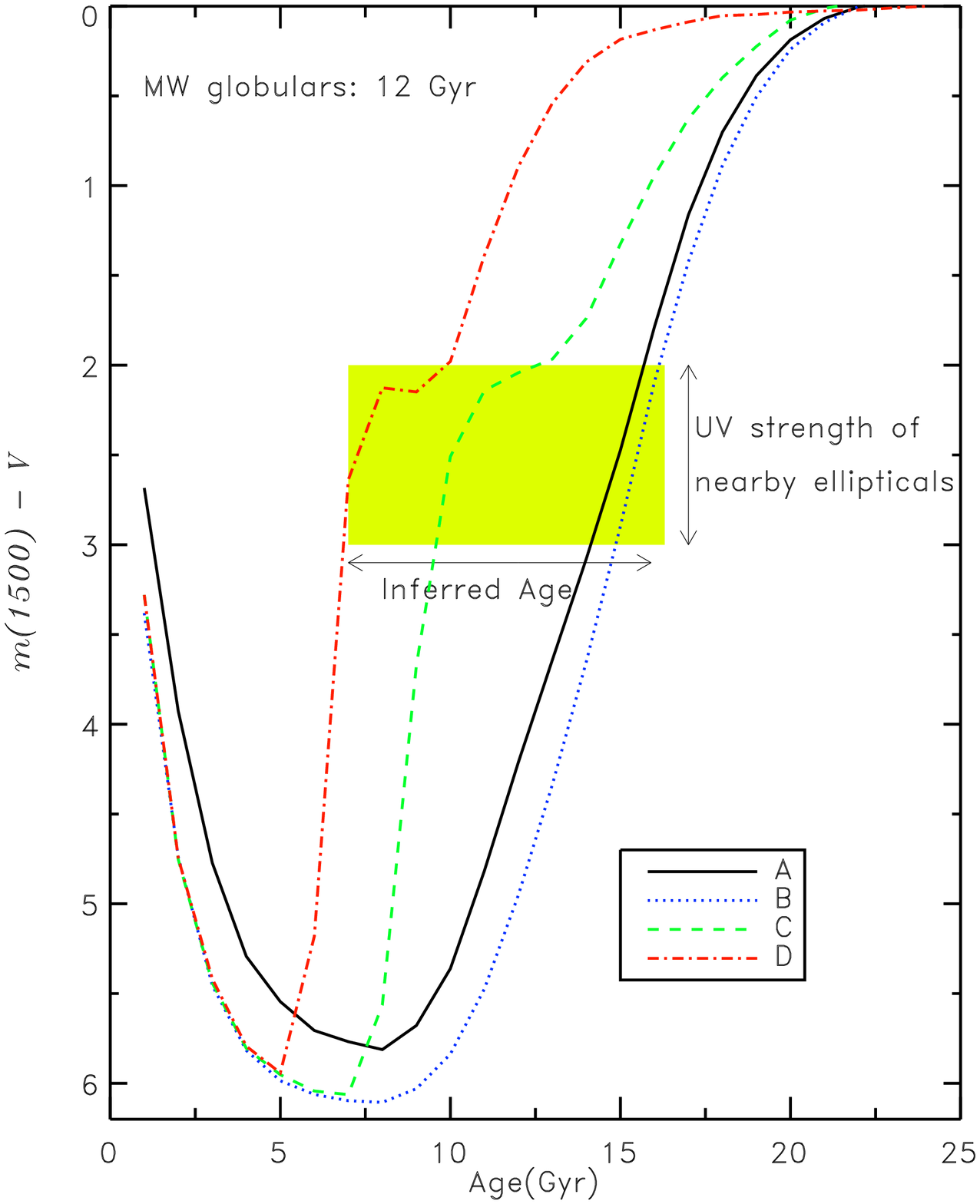]{
The evolution of the UV-to-$V$ flux ratio as a function of time.
One magnitude difference means about a 2.5 times difference in the mean flux.
All models are capable of matching the observed flux ratio (from Dorman et al.
1995) and have similar trends in developing a high UV flux. However, they
infer the ages for giant ellipticals that are different by a factor of two.
\label{fig7}}

\figcaption[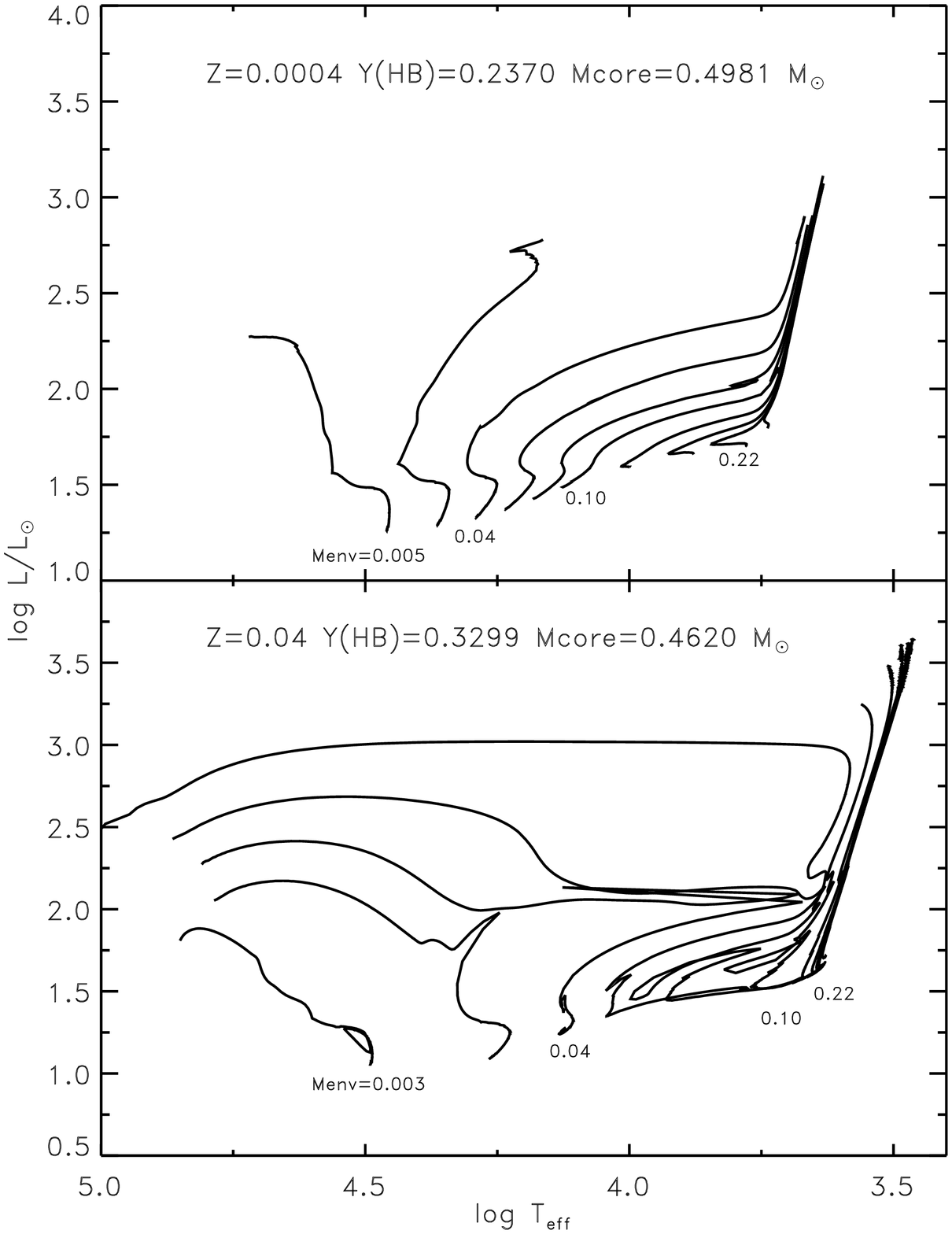]{
The HB evolutionary tracks for metal-poor ($top$) and metal-rich ($bottom$) 
stars. For the same metallicity, stars are assumed to have the same core mass.
Some tracks are accompanied by the envelope mass in \Ms.
Note that the temperature of the HB stars gradually increase as their masses
decrease in the metal-poor population, while it is much more abrupt
in the metal-rich case.
\label{fig8}}

\figcaption[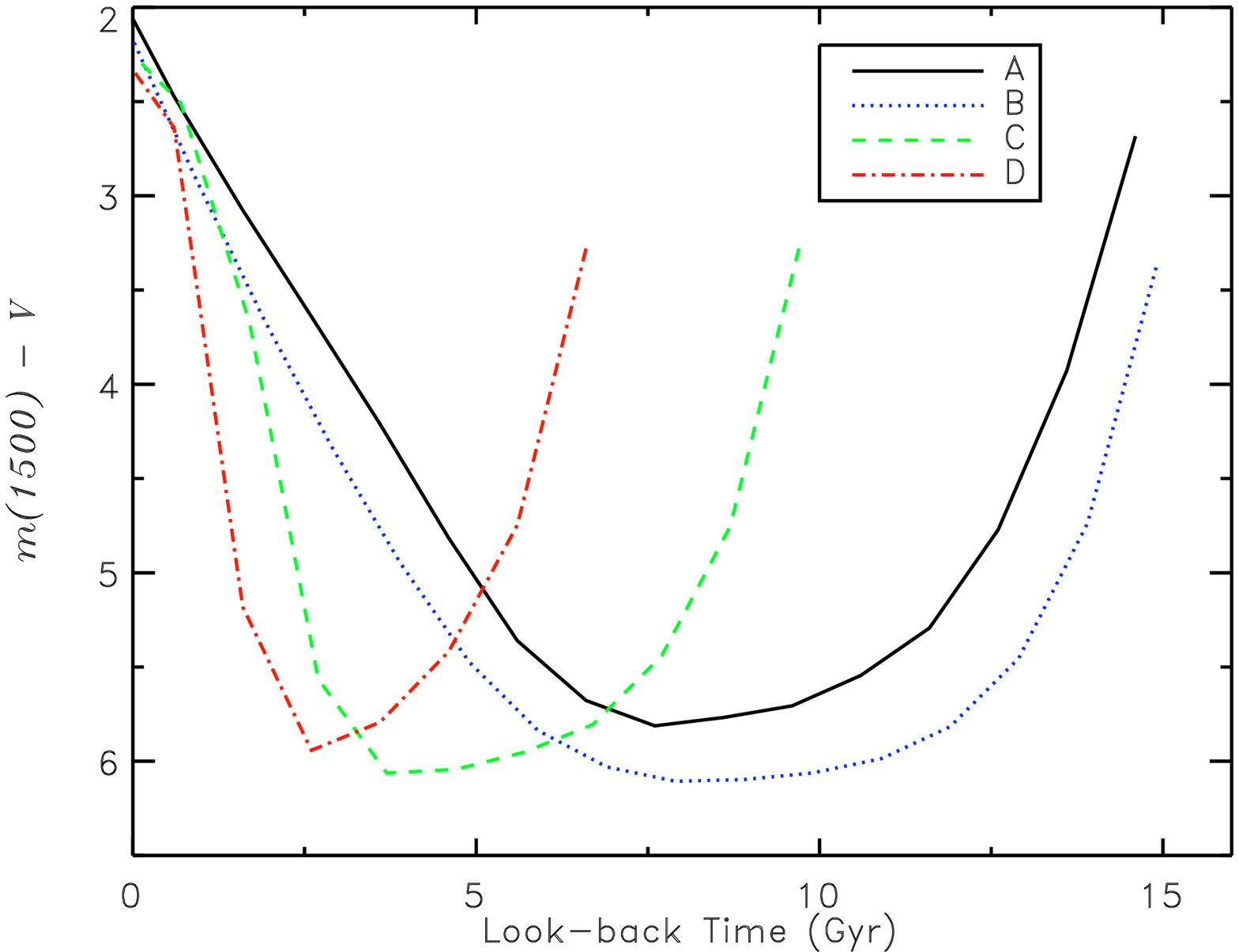]{
The UV-to-$V$ flux ratio as a function of lookback time.
Each model is based on the assumption that old giant elliptical galaxies at 
present epoch are as old as their UV-to-$V$ flux ratios suggest, as shown in 
Table 3.
\label{fig9}}

\figcaption[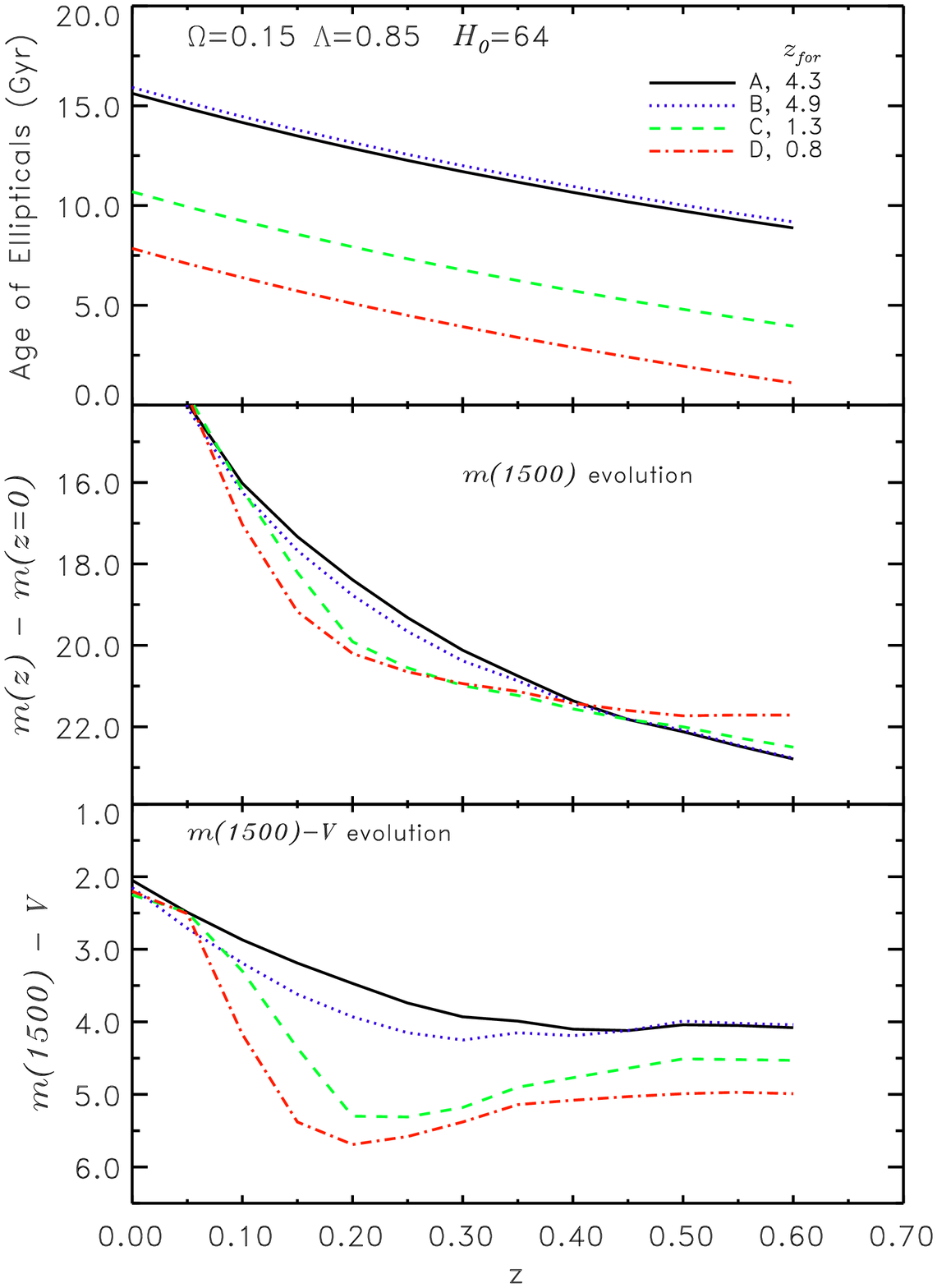]{
The evolution of the UV upturn as a function of redshift.
$Top$: the predicted ages of giant ellipticals as functions of redshift. 
The predicted UV magnitude ($middle$) and the UV-to-$V$ flux ratio ($bottom$)
in observer's magnitude. Note that the younger models (C \& D) fade
in the UV much faster as redshift increases (middle panel).
As a result, a large difference in the UV-to-$V$ flux ratio between
different models is predicted.
\label{fig10}}

\figcaption[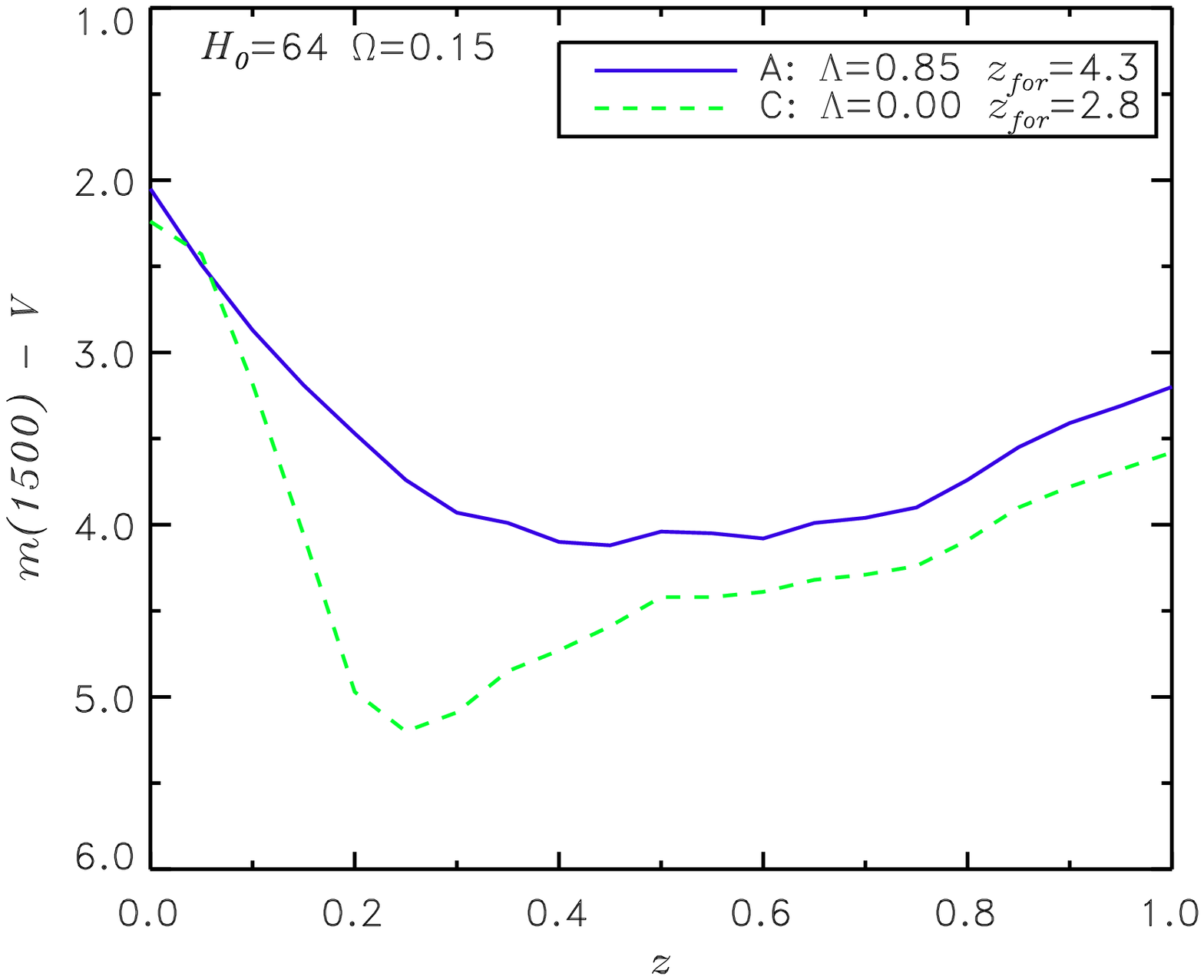]{
The two representative models (Model A: ``old'', Model C: ``young'') can
tell us about the mean age of giant elliptical galaxies with strong UV
upturns, regardless of the adopted cosmology. Models are displayed
according to different cosmology models from Fig. 10, yet, a similarly large
difference between Models A and C is present. So, the UV upturn test, as
a means of exploring the ages of galaxies is insensitive to cosmology.
\label{fig11}}

\clearpage

\begin{table*}
\caption{Choice of the mass loss parameter $\eta$.} \label{tbl-1}
\begin{center}
\begin{tabular}{cccc}
\tableline
\tableline
$\Delta$M & Mean Age of GCs &  Mean Z of GCs & $\eta$ \\
\tableline
0.257 \Ms  &    12 Gyr&     0.001 &      0.7\\
0.208 \Ms  &    15 Gyr&     0.001 &      0.5\\
\tableline
\tableline
\end{tabular}
\end{center}
\end{table*}

\begin{table*}
\caption{The variable-$\eta$ hypothesis\tablenotemark{\dagger}.}\label{tbl-2}
\begin{center}
\begin{tabular}{cc}
\tableline
\tableline
$Z$  & suggested $\eta$ \\
\tableline
$0.0001$ & 0.5  \\
$0.0004$ & 0.5  \\
$0.001$  & 0.7  \\
$0.004$  & 0.7  \\
$0.01$   & 0.7  \\
$\gtrsim 0.02$  & 1.0  \\
\tableline
\tableline
\end{tabular}
\end{center}
\tablenotetext{\dagger}{Same as in Yi et al. (1998), except that
$Z$ = 0.0001 and 0.001 have been newly added in this study.}
\end{table*}

\begin{table*}
\caption{Population synthesis model descriptions.} \label{tbl-3}
\begin{center}
\begin{tabular}{cccccccc}
\tableline
\tableline
Model & Chem. Evol. & \DYDZ & $\eta$  & Age(Gyr)\tablenotemark{\dagger}&major UV sources\\
\tableline
A     & simple      & 2.0   & 0.7     & 15.4    &meta-poor \\
B     & infall      & 2.0   & 0.7     & 15.9    &both metal-rich \& poor\\
C     & infall      & 2.5   & 0.7     & 10.7    &metal-rich \\
D     & infall      & 2.5   & variable&  7.6    &metal-rich \\
\tableline
\tableline
\end{tabular}
\end{center}
\tablenotetext{\dagger}{The inferred age of nearby giant elliptical galaxies, assuming old Galactic globular clusters are approximately 12 Gyr old.}
\end{table*}

\newpage

\plotone{f1.eps}
\clearpage
\plotone{f2.eps}
\clearpage
\plotone{f3.eps}
\clearpage
\plotone{f4.eps}
\clearpage
\plotone{f5.eps}
\clearpage
\plotone{f6.eps}
\clearpage
\plotone{f7.eps}
\clearpage
\plotone{f8.eps}
\clearpage
\plotone{f9.eps}
\clearpage
\plotone{f10.eps}
\clearpage
\plotone{f11.eps}


\newpage

\begin{thebibliography}{}
\bibitem[Alcock et al. 1997]{a97}  Alcock, C. et al. 1997, \apj, 482, 89
\bibitem[Allen 1976]{a76}  Allen, C. W. 1976, in Astrophysical Quantities (The Athlone Press: London), 202
\bibitem[Arimoto 1996]{a96} Arimoto, N. 1996, in ASP Series 98, From Stars to Galaxies, ed. C. Leitherer, U. Fritze-v. Alvensleben, \& J. Huchra (ASP), 287
\bibitem[Audouze \& Tinsley 1976]{at76} Audouze, J., \& Tinsley, B. M. 1976, \araa, 14, 43
\bibitem[Bessell 1990]{bes90} Bessell, M. S. 1990, \pasp, 102, 1181
\bibitem[Bessell et al. 1997]{bcp98} Bessell, M. S., Castelli, F., \& Plez, B. 1998, \aap, 333, 231
\bibitem[Bica 1988]{bica88} Bica, E. 1988, \aap, 195,76
\bibitem[Bl\"{o}cker \& Sch\"{o}nberner 1990]{bs90} Bl\"{o}cker, T., \& Sch\"{o}nberner, D. 1990, \aap, 240, L11
\bibitem[Bowen \& Willson 1991]{bw91} Bowen, G. H., \& Willson, L. A. 1991, \apj, 371, L53
\bibitem[Bower, Lucey, \& Ellis 1992]{ble92} Bower, R. G., Lucey, J. R., \& Ellis, R. S. 1992, \mnras, 254, 601
\bibitem[Bressan et al. 1994]{bcf94}  Bressan, A., Chiosi, C., \& Fagotto, F. 1994, \apjsupp, 94, 63
\bibitem[Brown et al. 1995]{bfd95}  Brown, T. M., Ferguson, H. C., \& Davidsen, A. F. 1995, \apj, 454, L15
\bibitem[Brown et al. 1997]{bfdd97}  Brown, T. M., Ferguson, H. C., Davidsen, A. F., \& Dorman, B. 1997, \apj, 482, 685
\bibitem[Brown et al. 1998]{bro98}  Brown, T. M., Ferguson, H. C., Stanford, S. A., \& Deharveng, J.-M. 1998, \apj, in press
\bibitem[Burstein et al. 1988]{b88}  Burstein, D., Bertola, F., Buson, L. M., Faber, S. M., \& Lauer, T. R. 1988, \apj, 328, 440
\bibitem[Carroll et al. 1992]{cpt92} Carroll, S. M., Press, W. H., \& Turner, E. L. 1992, \araa, 30, 499
\bibitem[Castellani \& Tornamb\'{e} 1991]{ct91}  Castellani, M., \& Tornamb\'{e}, A. 1991, \apj, 381, 393
\bibitem[Chaboyer et al. 1996]{cds96}  Chaboyer, B., Demarque, P., \& Sarajedini, A. 1996, \apj, 459, 558
\bibitem[Chaboyer et al. 1998]{cha98}  Chaboyer, B., Demarque, P., Kernan, P. J., \& Krauss, L. M. 1998, \apj, 494, 96
\bibitem[Chiosi 1980]{c80} Chiosi, C. 1980, \aap, 83, 206
\bibitem[Chiosi et al. 1997]{cvb97} Chiosi, C., Vallenari, A., \& Bressan, A. 1997, A\&A Suppl., 121,301
\bibitem[Cole \& Deupree 1980]{cd80}  Cole, P. W., \& Deupree, R. G. 1980, \apj, 239, 284
\bibitem[Cole et al. 1985]{cdd85}  Cole, P. W., Demarque, P. \& Deupree, R. G. 1985, \apj, 291, 291
\bibitem[Demarque et al. 1996]{d96} Demarque, P., Chaboyer, B., Guenther, D., Pinsonneault, L., Pinsonneault, M., \& Yi, S. 1996, Yale Isochrones 1996 (available through Sukyoung Yi) 
\bibitem[Dorman et al. 1993]{dro93}  Dorman, B., Rood, R. T., \& O'Connell, R. 1993, \apj, 419, 596
\bibitem[Dorman et al. 1995]{dor95}  Dorman, B., O'Connell, R., \& Rood, R. T. 1995, \apj, 442, 105
\bibitem[Dunlop et al. 1996]{dun96} Dunlop, J., Peacock, J., Spinrad, H., Dey, A., Jimenez, R., Stern, D., \& Windhorst, R. 1996, Nature, 381, 481
\bibitem[de Boer 1987]{deb87} de Boer, K. 1987, in The Second Conference on Faint Blue Stars, eds. A. G. Davis Philip, D. S. Hayes, and J. W. Liebert (L. Davis Press: New York), 95
\bibitem[Feast \& Catchpole 1997]{fc97} Feast, M. W., \& Catchpole R. M. 1997, \mnras, 286, L1
\bibitem[Gibson \& Matteucci 1997]{gm97} Gibson, B. K., \& Matteucci, F. 1997, \mnras, 291, L8
\bibitem[Gratton et al. 1997]{gra97}  Gratton, R. G., Fusi Pecci, F., Carretta, E., Clementini, G., Corsi, C. E., \& Lattanzi, M. G. 1997, \apj, 491, 749
\bibitem[Greggio \& Renzini 1990]{gr90} Greggio, L., \& Renzini, A. 1990, \apj, 364, 35
\bibitem[Guenther \& Demarque 1997]{gd97} Guenther, D. B., \& Demarque, P. 1997, \apj, 484, 937
\bibitem[Guiderdoni \& Rocca-Volmerange 1987]{grv87} Guiderdoni, B., \& Rocca-Volmerange, B. 1987, \aap, 186, 1
\bibitem[Guth 1981]{g81} Guth, A. H. 1981, Phys. Rev. D, 23, 347
\bibitem[Heap et al. 1998]{hea98} Heap, S. R. et al. 1998, \apj, 492, L131
\bibitem[Heber et al. 1984]{heb84} Heber, U., Hunger, K., Jonas, G., \& Kudritzki, R. P. 1984, \aap, 130, 119
\bibitem[Hill 1982]{h82} Hill, G. 1982, Publications of the Dominion Astrophysical Observatory, 16-6, 67
\bibitem[Horch et al. 1992]{hdp92}  Horch, E., Demarque, P., \& Pinsonneault, M. 1992, \apj, 388, L53
\bibitem[Iben \& Renzini 1983]{ir83} Iben, I. Jr., \& Renzini, A. 1983, \araa, 21, 271
\bibitem[Izotov et al. 1997]{itl97} Izotov, Y. I., Thuan, T. X., \& Lipovetsky, V. A. 1997, \apjsupp, 108, 1
\bibitem[Kochanek 1996]{k96}  Kochanek, C. S. 1996, \apj, 466, 638
\bibitem[Kodama \& Arimoto 1997]{ka97}  Kodama, T., \& Arimoto, N. 1997, \aap, 320, 41
\bibitem[Kurucz 1992]{k92} Kurucz, R. 1992, in The Stellar Population in Galaxies, ed. B. Barbuy \& A. Renzini (Dordrecht: Reidel), 225
\bibitem[Landsman et al. 1996]{lan96}  Landsman, W. B., Sweigart, A. V., Bohlin, R. C., Neff, S. G., O'Connell, R. W., Roberts, M. S., Smith, A. M., \& Stecher, T. P. 1996, \apj, 472, L93
\bibitem[Larson 1972]{l72} Larson, R. B. 1972, Nature Physical Science, 236, 62
\bibitem[Lee 1990]{l90}  Lee, Y.-W. 1990, \apj, 363, 159
\bibitem[Lee 1994]{l94}  Lee, Y.-W. 1994, \apj, 430, L113
\bibitem[Lee et al. 1990]{ldz90}  Lee, Y.-W., Demarque, P., \& Zinn, R. 1990, \apj, 350, 155
\bibitem[Lee et al. 1994]{ldz94}  Lee, Y.-W., Demarque, P., \& Zinn, R. 1994, \apj, 423, 248
\bibitem[Lequeux et al. 1979]{l79} Lequeux, J., Peimbert, M., Rayo, J. F., Serrano, A., \& Torres-Peimbert, S. 1979, \aap, 80, 155
\bibitem[Maeder 1992]{m92} Maeder, A. 1992, \aap, 264, 105
\bibitem[Martin et al. 1998]{m98} Martin, C. et al. 1998, BAAS, 29, 1309 
\bibitem[Michaud et al. 1983]{mvv83} Michaud, G., Vauclair, G., \& Vauclair, S. 1983, \apj, 267, 256
\bibitem[Minniti 1995]{m95} Minniti, D. 1995, \aj, 109, 1663
\bibitem[Morossi et al. 1993]{m93} Morossi, C., Franchini, M., Malagnini, M. L.,  Kurucz, R. L., \& Buser, R. 1993, \aap, 277, 173
\bibitem[Pagel et al. 1992]{pag92} Pagel, B. E. J., Simonson, E. A., Terlevich, R. J., \& Edmunds, M. G. 1992, \mnras, 255, 325
\bibitem[Park \& Lee 1997]{pl97}  Park, J.-H., \& Lee, Y.-W. 1997, 476, 28
\bibitem[Peacock et al. 1998]{pea98} Peacock, J. A., Jimenez, R., Dunlop, J. S., Waddington, I., Spinrad, H., Stern, D., Dey, A., \& Windhorst, R. A. 1998, \mnras, 296, 1089
\bibitem[Peimbert 1995]{p95} Peimbert, M. 1995, in The Light Element Abundances, ed. P. Crane (Springer-Verlag), 165
\bibitem[Reid 1997]{r97} Reid, I. N. 1997, \aj, 114, 161
\bibitem[Reimers 1975]{r75} Reimers, D. 1975, M\'{e}m. Soc. Roy. Sci. Li\`{e}ge, 6th Ser., 8, 369
\bibitem[Reiss et al. 1996]{rpk96} Reiss, A. G., Press, W. H., \& Kirshner, R. P. 1996, \apj, 473, 88
\bibitem[Riess et al. 1998]{r98} Riess, S. et al. 1998, \aj, in press
\bibitem[Salaris \& Weiss 1997]{sw97} Salaris, M., \& Weiss, A. 1997, \aap, 327, 107
\bibitem[Sch\"{o}nberner 1979]{s79}  Sch\"{o}nberner, D. 1979, \aap, 79, 108
\bibitem[Sch\"{o}nberner 1983]{s83}  Sch\"{o}nberner, D. 1983, \apj, 272, 708
\bibitem[Sosin et al. 1997]{sos97}  Sosin, C., Dorman, B, Djorgovski, S. G., Piotto, G., Rich, R. M., King, I. R., Liebert, J., Phinney, E. S., \& Renzini, A. 1997, \apj, 480, L35
\bibitem[Spinrad et al. 1997]{spi97} Spinrad, H., Dey, A., Stern, D., Dunlop, J., Peacock, J., Jimenez, R., \& Windhorst, R. 1997, \apj, 484, 581
\bibitem[Sweigart 1998]{swe98} Sweigart, A. V. 1998, in The Third Conference on Faint Blue Stars, eds. A. G. Davis Philip (New York: L. Davis Press), in press
\bibitem[Tantalo et al. 1996]{tan96} Tantalo, R., Chiosi, C., Bressan, A., \& Fagotto, F. 1996, \aap, 311, 361
\bibitem[Tomasko 1970]{t70} Tomasko, M. G. 1970, \apj, 162, 125
\bibitem[Trimble 1987]{t87} Trimble, V. 1987, \araa, 25, 425
\bibitem[VandenBerg et al. 1996]{vbs96} VandenBerg, D. A., Bolte, M., \& Stetson, P. B. 1996, \araa, 34, 461
\bibitem[Weidemann \& Koester 1983]{wk83} Weidemann, V., \& Koester, D. 1983, \aap, 121, 77
\bibitem[Weidemann 1997]{wei97} Weidemann, V. 1997, in Advances in Stellar Evolution, eds. R. T. Rood and A. Renzini (Cambridge University Press: Cambridge), 169
\bibitem[Willson et al. 1996]{wbs96} Willson, L. A., Bowen, G. H., \& Struck, C. 1996, in ASP Series 98, From Stars to Galaxies, eds. C. Leitherer, U. Fritze-v. Alvensleben, \& J. Huchra (ASP), 197
\bibitem[Yi et al. 1995]{yado95}  Yi, S., Afshari, E., Demarque, P., \& Oemler, A. Jr. 1995, \apj, 453, L69
\bibitem[Yi et al. 1997a]{ydk97}  Yi, S., Demarque, P., \& Kim, Y.-C. 1997a, \apj, 482, 677
\bibitem[Yi et al. 1997b]{ydo97}  Yi, S., Demarque, P., \& Oemler, A. Jr. 1997b, \apj, 486, 201
\bibitem[Yi et al. 1998]{ydo98}  Yi, S., Demarque, P., \& Oemler, A. Jr. 1998, \apj, 492, 480

\end{thebibliography}
\end{document}